# Is there Lower Limit to Velocity or Velocity Change?

*B N Sreenath*

S N Bhat P U College of Science & Commerce, Bangalore, India

e-mail: bnsreenath@yahoo.co.in

*Kenath Arun*

Christ Junior College, Bangalore, 560 029, India

Telephone: +91-80-4012 9292; Fax: +91-80- 4012 9222

e-mail: kenath.arun@cjc.christcollege.edu

*C Sivaram*

Indian Institute of Astrophysics, Bangalore, 560 034, India

Telephone: +91-80-2553 0672; Fax: +91-80-2553 4043

e-mail: sivaram@iiap.res.in

**Abstract:** Here we explore the possibility of a lower limit to velocity or velocity change which is 20 orders of magnitude smaller than the speed of light and explore the various observational signatures.

Special relativity implies an upper limit to velocity, i.e. that of light propagating in vacuum. This is well tested. Recent gamma ray bursts (Sivaram, 2000; Tanvir, 2009) with afterglows have enabled the independence of this upper limiting velocity with frequency in different parts of the electromagnetic spectrum from radio waves to gamma rays, and also in gamma rays of different energies. The fractional deviation (limit) is expressed as:

$$\frac{\Delta c}{c} < 10^{-18} \qquad \ldots (1)$$

However special relativity is strictly valid only in inertial frames (in Euclidean flat space or absence of gravitational fields). Gravity by bending light causes light to propagate at lower speed (like a refracting medium) (Misner, Thorne & Wheeler, 1973). It is not a priori evident that c should also be the limiting velocity in microscopic (subatomic) regions. Indeed for the highest energy cosmic rays observed (~$10^{20}$eV), the deviation from light velocity is ~$10^{-20}$.

In the micro-world (on the contrary) one would expect, especially around fermi length scales or below, the existence of a lower limit to speed of propagation. These have been discussed in different contexts (deformed space-time, special relativity formulated with a minimal length scale, etc.) (Sivaram, 1993; Kloznaik, 1999; Lukierski et al, 1995). Recently a suggestion (Sreenath, 2012) was made that there could be a minimal velocity and its value was mentioned. Now arguing from the wavelength of de Broglie waves, i.e.:

$$\lambda = \frac{h}{mv} \quad (\text{say v} \sim c)$$

then for a change in velocity, we have:

$$d\lambda = \frac{h}{mv^2} dv \qquad \ldots (2)$$

Now from various considerations, we may expect SR to become invalidated, at the Planck scale $\sim \left(\frac{\hbar G}{c^3}\right)^{1/2} \sim 10^{-33} cm$, when gravity and curvature effects become locally very important!

Thus the requirement that

$$d\lambda \geq \left(\frac{\hbar G}{c^3}\right)^{1/2} \qquad \ldots (3)$$

implies from equation (2), that (for m ~ $m_e$ of electron mass)

$$dv \geq 10^{-11} cm/s \qquad ...(4)$$

So this gives a lower limit to change in propagation velocity:

$$dv \approx m_e c^2 \left(\frac{G}{\hbar c^3}\right)^{1/2} \approx 10^{-11} cm/s \qquad ...(5)$$

This is the same as the value obtained in (Sreenath, 2012), under very different considerations.

Again the smallest energy associated in the universe is in cosmology, i.e. $E_{min} \approx \hbar H_0$, where $H_0$, the Hubble constant gives the smallest frequency of $10^{-18}$Hz, implying $E_{min} \approx 10^{-45} ergs$. (Sivaram, 1986a)

This minimum energy has also been identified with the quantum of mass in cosmology, giving the universality of the self-gravitational potential energy for a fundamental particle (Alfonso-Faus, 2012). Such a significance for the Weinberg relation as a minimum measurable gravitational potential energy was also pointed out earlier (Sivaram, 1982; 1983). This was also shown to lead to a cosmological constraint on Hawking black hole temperatures (Sivaram, 1983).

A proton if it acquires this kinetic energy (at almost zero velocity), would have a velocity again given as:

$$v_{min} \approx \left(\frac{\hbar H_0}{m_p}\right)^{1/2} \approx 10^{-11} cm/s \qquad ...(6)$$

Same as in the above relation!

The Weinberg relation and the minimum quantum gravitational energy was reinterpreted (Sivaram, 1994) in terms of the dominant cosmological constant (accounting for 0.8 of critical density) as:

$$E_{min} = \hbar c \sqrt{\Lambda} \qquad ...(7)$$

(if as current observations indicate, a constant $\Lambda$ is nothing but the dark energy, this has more fundamental significance!)

This then gives a minimum velocity of:

$$v_{min} = \left(\frac{\hbar c \sqrt{\Lambda}}{m_P}\right)^{1/2} \approx 10^{-11} cm/s \qquad \text{... (8)}$$

(for the observed $\Lambda \approx 10^{-56} cm^{-2}$)

It was also pointed out (Sivaram, 1994) that $\Lambda$ also implies the minimal MOND acceleration of:

$$a_{min} = c^2 \sqrt{\Lambda} \approx 10^{-8} cm/s^2 \qquad \text{... (9)}$$

Thus both $v_{min}$ and $a_{min}$ may have a fundamental cosmological basis.

Moreover for high energy cosmic ray particles (say protons) of energy E we have: (Sivaram, 1986a; 1986b)

$$\frac{c-v}{c} \approx \left(\frac{m_0 c^2}{E}\right)^2 \approx 10^{-22} \qquad \text{... (10)}$$

(for $E \sim 10^{20} eV$)

This again implies $(c-v) \approx 10^{-11} cm/s$.

Gamma ray bursts are not yet sensitive to $\frac{\Delta c}{c} \approx 10^{-22}$. If they are detected at z ~20, we can approach this limit.

All these limits arrived from entirely different phenomena, possibly point to the existence of a lower limiting speed of $\sim 10^{-11} cm/s$. This is testable in future experiments (and may also explain why cosmic rays above $3 \times 10^{20} eV$ have not been detected even after the 'falling brick' event of above energy was seen two decades ago!).

Instances where such a lower limiting speed could manifest are in atomic spectroscopy where the corrections to the wavelengths of standard lines (especially to the hyperfine transitions like Tritium which are known to 17 decimal places) would be expected to change by one part in $10^{21}$. And again gyro-frequencies (e.g. of protons in low magnetic fields) could change by the same amount. These effects are testable in future high precision experiments. This existence of such a lower limiting velocity is not excluded by any principle or law in physics.

Also it has been suggested that in a large class of quantum gravity approaches (Sivaram, 2000); a deformed photon dispersion arises which is of the form:

$$Pc = E\left(1 + \frac{E}{E_{QG}}\right)^{1/2} \qquad \text{... (11)}$$

Where $E_{QG}$ is the quantum gravity scale.

One of the consequences could be that high energy photons would not travel at the speed of light, but at a speed of:

$$v = c\left(1 - \frac{E}{E_{QG}}\right) \qquad \text{... (12)}$$

Thus for $\Delta t$ we have:

$$\Delta t = \frac{D}{c}\left(\frac{E}{E_{QG}}\right) \qquad \text{... (13)}$$

Here:

$$\Delta t = \frac{D}{c} - \frac{D}{c + \Delta c} = \frac{\Delta c}{c}\frac{D}{c} \qquad \text{... (14)}$$

So for $\frac{\Delta c}{c} \approx 10^{-22}$ (as required in the case of a minimal change in velocity): $\Delta t = 10^{-22}\frac{D}{c}$

So for a source at $3 \times 10^{28}$ cm, this implies: $\Delta t \approx 10^{-4} s$

That is photons of different energies should be received within a millisecond. This implies an equivalent quantum gravity scale of $\sim 10^{20} GeV$ (as $\Delta t \propto \frac{1}{E_{QG}}$ from equation (13))

Again the lowest temperature, one could conceive of, would correspond to Hawking temperature of a black hole (with a mass that of the universe) which is $\sim 10^{-29}$ degrees.

This is also Unruh temperature associated with the minimal MOND acceleration of $10^{-8}$ cm/s$^2$, i.e. $\frac{\hbar a}{2\pi k_B} \sim 10^{-29}$. The corresponding $kT \sim 10^{-45} ergs$ giving again a minimal velocity of $10^{-11} cm/s$.

Corresponding to the well discussed MOND acceleration (Sivaram, 1994); which modifies Newtonian dynamics below this value, there could be a special relativistic modification at this minimal velocity of $10^{-11} cm/s$. So from a number of considerations in widely different contexts we can suggest such a minimal velocity. We have also suggested possible signatures.